\documentstyle[prb,aps,multicol]{revtex}

\newcommand{\be}{\begin{equation}}
\newcommand{\ee}{\end{equation}}
\newcommand{\bea}{\begin{eqnarray}}             
\newcommand{\eea}{\end{eqnarray}}
\newcommand{\nn}{\nonumber}
\newcommand{\bm}[1]{\mbox{\boldmath${#1}$}}
\newcommand{\grad}{\bm{\nabla}}

\begin{document}

\title{Scattering of Phonons by a Vortex in a Superfluid}       
\author{C.\ Wexler}      
\address{Department of Physics, Box 118440, 
         University of Florida,
         Gainesville, FL 32611-8440}
\author{D.\ J.\ Thouless}
\address{Department of Physics, Box 351560, 
         University of Washington, 
         Seattle, WA 98195-1560}
\date{August 10$^{\rm th}$, 1998}
\draft
\maketitle

\begin{abstract}
Recent work gives a transverse force on an isolated moving vortex
which is independent of the normal fluid velocity, but it is widely
believed that the asymmetry of phonon scattering by a vortex leads to
a transverse force dependent on the relative motion of the normal
component and the vortex.  We show that a widely accepted derivation
of the transverse force is in error, and that a careful evaluation
leads to a much smaller transverse force.  We argue that a different
approach is needed to get the correct expression.
\end{abstract}
\pacs{67.40.Vs,67.57.Fg,47.37.+q,47.32.Cc}

\begin{multicols}{2}

\section{Introduction}
\label{sec:introduction}

Quantized vortices have been an essential part of the theory of
superfluids for the best part of the last few decades
\cite{vortices}. Vortices provide an important mechanism for
the decay of supercurrents in either superfluids and superconductors,
and their motion is responsible for dissipative effects.
A considerable effort has been devoted to the derivation of the
equation of motion for vortices. 

It is a widely held concept that for a vortex moving with velocity
${\bf v}_V$ through a superfluid there is a force transverse to its
velocity in analogy to the Magnus or Kutta-Joukowski force in
classical hydrodynamics \cite{lamb}. This {\em transverse force} is given
by the usual expression  
\be
  \label{eq:magnus}
  {\bf F}_{\rm t} = \rho_s \; \bm{\kappa}_0 \times 
        ( {\bf v}_V - {\bf v}_s ) \;,
\ee
where $\rho_s$ is the density of the superfluid component,  
$\bm{\kappa}_0$ is a vector parallel to the vortex whose magnitude is
equal to the quantum of circulation $h/m$, and ${\bf v}_s$ is the 
superfluid velocity in the vicinity of the vortex. At zero temperature
the superfluid density equals the total fluid density and this force
is all there is to the story. At finite temperatures, however, part of
the fluid is in the {\em normal component} and there is the
possibility of additional transverse forces proportional to the
velocity difference relative to this normal component, 
$({\bf v}_V - {\bf v}_n)$. The existence and value of this additional
transverse force has been quite controversial and very different
expressions for it can be found in the literature 
\cite{volovik,TAN,wexler97,review,barenghi,sonin97}. 

Recent work by Thouless, Ao and Niu \cite{TAN} and Wexler
\cite{wexler97}, unambiguously shows that the transverse
force per unit length on a single quantized vortex in a superfluid is
given by the universal expression in eq.\ \ref{eq:magnus}, without 
any additional transverse forces. The argument is based on a global
argument that relies on the equilibration of the excitations, so it
does not deal separately with the change in momentum due to
phonon-vortex scattering and the modification of the phonon-phonon
scattering due to the vortex motion.
In essence the argument is as follows: Wexler's thermodynamic argument
\cite{wexler97} shows that the coefficient of 
$\bm{\kappa}_0 \times {\bf v}_s$ is indeed $-\rho_s$. This result does
not appear to be controversial. Thouless, Ao and Niu \cite{TAN} showed
that the part of the transverse that depends on the {\em vortex
velocity} ${\bf v}_V$ is given by 
\be
\label{eq:circulations}
        (\rho_s \; \bm{\kappa}_s + \rho_n \; \bm{\kappa}_n ) 
        \times {\bf v}_V \;,
\ee
where $\rho_n$ is the density of the normal fluid component and
${\kappa}_s$, ${\kappa}_n$ are the {\em equilibrium} circulations of
the superfluid and normal fluid parts far away from the vortex
core. The circulation of the superfluid is clearly the quantum of
circulation $\kappa_0 = h/m$ and the normal fluid is assumed 
irrotational $\kappa_n = 0$ due to its viscosity \cite{circulation}.
Since the coefficients of ${\bf v}_V$ and ${\bf v}_s$ are equal and
opposite Galilean invariance leaves no room left for any additional
transverse force depending on the normal fluid velocity ${\bf v}_n$.

The arguments presented are rather straightforward but there is still
considerable belief in the existence of additional forces. One should
then consider the alternative calculations more closely and ask what
is the origin of this controversy. 
In part, the confusion arises from different interpretations of the
role played by the scattering of excitations the vortex. This is
discussed next.

\section{Scattering of excitations by a vortex}
\label{sec:cleary}

There is no doubt that scattering of excitations by a vortex produces
a longitudinal force proportional to the relative velocity of the
excitations and the vortex 
\cite{barenghi,sonin97,vinen,pitaevskii,fetter,iordanskii,demircan}. 
For either phonons or rotons this force is {\em quadratic} in the
circulation $\kappa$, and at temperatures low enough that phonons
become dominant is {\em fifth} power in temperature, if only
phonon-vortex scattering is taken into account and the perturbation of
the phonon--phonon scattering due to the vortex motion is ignored. 

It is widely believed, however, that there is also a transverse force
due to an asymmetry in this phonon-vortex scattering.  The details of
this are not clear to us from Iordanskii's papers \cite{iordanskii},
but a  derivation by Cleary \cite{cleary} which gives a transverse 
force linear in $\kappa$ and fourth power in temperature (like the
normal fluid density due to phonons) is widely quoted, for example by
Sonin \cite{sonin97}. Demircan {\em et al.}\cite{demircan}, on the 
other hand, find that this force vanishes. 

It has been argued \cite{sonin97,shelankov} that the divergence of the
scattering amplitude at small angles invalidates arguments, such as
those in reference \onlinecite{demircan}, based on the Born
approximation for the scattering amplitude. 
In fact the $\theta^{-1}$ divergence of the scattering amplitude for small
scattering angles $\theta$ gives no trouble with the longitudinal force,
which depends on the scattering cross section weighted with
$(1-\cos\theta)$, and only a formal divergence for the transverse force,
where weighting of the differential cross section with $\sin\theta$ leads
to a principle part integral and the need for a regularization
procedure.

Examination of the Cleary derivation shows that the derivation is in error. 
Whereas Cleary obtained a transverse force per unit length proportional to 
\be
        \label{eq:cleary}
        (v_V-v_n)\kappa_0 T^4/\hbar^3c^5\;,
\ee
a correct evaluation of the formula he used for phonon-vortex scattering
gives an answer which term by term is of order
\be
        \label{eq:cleary'}
        (v_V-v_n)\kappa_0^3 T^6/\hbar^5c^9\;.
\ee
Cleary reached eq.\ \ref{eq:cleary} by adding and subtracting divergent
series to his expression. Demircan et al. did not get eq.\
\ref{eq:cleary'} because they did not keep enough terms.  They did not
get eq.\ \ref{eq:cleary} because it does not follow from Cleary's
starting-point. 

Let us consider a phonon in a superfluid with an infinite rectilinear
vortex on the $z$-axis. We assume a stationary vortex for the
calculations that follow. Galilean invariance can be used later to 
calculate the force in a more general situation \cite{wexler97} and the
oscillatory motion of the vortex due to the interaction with the
phonon is not important for the determination of the transverse force
(see  Sec.\ III D in Ref.\ \onlinecite{sonin97}).
Phonons interact with the gauge-like
superfluid velocity field and with the density variations near the
vortex core. In the long-wavelength limit the former dominates
\cite{fetter} and phonons satisfy the following equation of motion 
\be
\label{eq:phonon_eom}
        \frac{\partial^2 \psi}{\partial t^2} - c^2 \nabla^2 \psi 
        + 2 \; {\bf v}_{s0} (r) \cdot 
        \grad \frac{\partial \psi}{\partial t} = 0  \; ,
\ee
where $\psi$ represents the phonon wavefunction and 
${\bf v}_{s0} (r) = {\bf e}_\theta \kappa_0/(2 \pi r)$ is the
superfluid velocity around a vortex with circulation $\kappa_0$.
Given the symmetry of the problem and the simple form of eq.\
\ref{eq:phonon_eom}, it is possible to calculate the differential 
cross section for a phonon-vortex event exactly by using a 
partial-wave decomposition.
\be
        \psi ({\bf r}, t) = \sum_{l=-\infty}^{\infty} \sum_{k_z} 
        \psi_l (r) e^{i k_z z + i l \theta - i \omega t} \; ,
\ee
where $\omega = c |k| = c \sqrt{k_r^2 + k_z^2}$. Each partial-wave
satisfies (for $l \neq 0$)
\be
        \frac{d^2 \psi_l}{d r^2} + \frac{1}{r} \frac{d \psi_l}{d r} -
        \frac{(l + k \kappa_0/2 \pi c)^2}{r^2} \psi_l + 
        k_r^2 \psi_l = 0 \;,
\ee
where a small term proportional to ${v}_{s0}^2$ has been dropped
\cite{note}.
The solutions can be obtained in terms of ordinary Bessel functions
$J_{\nu_l} ( k_r r )$, where $\nu_l = | l + k \kappa_0/2 \pi c|$ for
$l \neq 0$ (note that from eq.\ \ref{eq:phonon_eom}, $\nu_0 = 0$ but
since s-waves have considerable overlap with the vortex core, in
general one should keep this case apart). One should also note that
this is not a Born approximation solution but rather the exact
solution of the phonon wave equation.

Upon consideration of the asymptotic behavior of the Bessel function
one can readily obtain the scattering amplitude 
\be
\label{eq:f}
        f (\theta) = \sqrt{\frac{2}{\pi k_r}}
                \sum_{l=-\infty}^{\infty} 
                e^{i \delta_l + i l \theta} 
                \sin \delta_l \; ,
\ee
where 
\be
\label{eq:delta_l}
\delta_l = \frac{\pi}{2} ( |l| - \nu_l ) = \eta \; {\rm sign} (l) \;,
\ee
and $\eta = - { k \kappa_0}/{4 c}$. Note that for a vortex in
superfluid helium $|\eta| \ll 1$ even for the most energetic phonons. 

The series in eq.\ \ref{eq:f} are not convergent, since all the
terms are of the same magnitude, but when we regularize the series in
an obvious way, by writing the phase shifts as 
\be\label{eq:regular}
\delta_l = \eta \: {\rm   sign}(l)\: \exp(-\alpha |l|)\;,\ {\rm with}\
0 < \alpha \ll 1\;,
\ee 
we get the scattering amplitude as
\be\label{eq:regamp}
f_{\rm reg}(\theta)=i\sqrt{2\over\pi k_r}{\sin\eta\cos(\eta+\theta/2)
  \over \sin(\theta/2)}\;,
\ee
in the limit $\alpha \rightarrow 0$. Here the phase shift $\delta_0$
has been taken to be zero. This gives the differential scattering
cross section as 
\bea
\label{eq:diffsig}
\frac{d\sigma}{d\theta} &=& 
        \frac{2}{\pi k_r} \sin^2\eta \: [\cos^2\eta\: \cot^2(\theta/2)
\nn \\ 
&& -2 \sin\eta \: \cos\eta \: \cot(\theta/2) +\sin^2\eta] \;.
\eea
In this form the differential cross section looks innocuous, giving a
well-defined transport cross section in terms of the integral weighted
with $1-\cos\theta$, and a {\em transverse scattering cross section},
in terms of the integral weighted with $\sin\theta$, that is a
principal part integral.  As we will see, a different limiting process
can, however, give a different answer for the transverse scattering
cross section, since the limiting behavior at very small angles can
change the integral. 

If we calculate the transverse cross section directly
from eq.\ \ref{eq:f} this gives
\bea
\label{eq:sigma_perp_partial_1}
\sigma_\perp & = &
\int_0^{2 \pi} d \theta \; \sin \theta \; | f (\theta) |^2 
 \\
&=& \frac{2}{\pi k_r} \int \sin\theta\, d \theta \;  \sum_{l,l'} \:
e^{i (l-l') \theta} \: e^{i (\delta_l - \delta_{l'})} \:
\sin \delta_l \: \sin \delta_{l'}
\nn \\
&=& \frac{4}{k_r} \sum_l \sin \delta_l \: \sin \delta_{l+1} \;
        \sin (\delta_l - \delta_{l+1}) \;,
\label{eq:sigma_perp_partial}
\eea
where the sums and integral were commuted to obtain the last
expression. This last step is not trivial given the non-uniform
convergence of the series. If the expression in eq.\
\ref{eq:sigma_perp_partial} is taken literally, then, for
$\delta_0=0$, every term in this series vanishes.  However, if the
large angular momentum phase shifts are smoothly taken to zero, as in
eq.\ \ref{eq:regular}, then the sum in eq.\
\ref{eq:sigma_perp_partial} can be expressed in terms of an integral
over $\delta_l$:
\bea
\sigma_\perp &=& \frac{8}{k_r} \, \int_{\delta_\infty}^{\delta_1}
\sin^2\delta_l \; \sin d \delta_l 
\simeq \frac{8}{k_r} \, \int_{0}^{\eta}
\sin^2\delta_l \; d \delta_l \nn \\
&=& \frac{2}{ k_r}(2\eta -\sin 2 \eta ) 
\simeq \frac{8 \: \eta^3}{3 k_r} \;. 
\label{eq:sigma_perp_aa}
\eea
If the same integral (eq.\ \ref{eq:sigma_perp_partial_1})
is performed for the expression for the differential cross section
given in eq.\ \ref{eq:diffsig}, then we get 
\be
\sigma_\perp=-\frac{4}{k_r}\:\sin^2\eta\: \sin\,2\eta\;.
\label{eq:sigma_perp_bb}
\ee
These two expressions, \ref{eq:sigma_perp_aa} and
\ref{eq:sigma_perp_bb}, are both of order $\eta^3$, but differ both
in sign and magnitude of the coefficient.  Examination of the
regularization procedure shows that in the range of angles
$\alpha<<\theta<<\sqrt{\alpha}$ the even part of the scattering
amplitude is proportional to $\eta^2\alpha/\theta^2$, while the odd
part is proportional to $\eta/\theta$.  The squares of each of these
terms give  contributions to the transport cross section which
vanishes in the limit $\alpha\to0$, but the product gives a
contribution to the transverse cross section that is of order
$\eta^3$. Let us look at this in more detail. We start with eqs.\
\ref{eq:f} and \ref{eq:delta_l}, and assume the same regularization
procedure as above, expanding this time in powers of $\eta$ while
keeping $\alpha$ finite throughout the calculation:
\bea
\label{eq:f_reg2}
f (\theta) &=& 
        \sqrt{\frac{2}{\pi k_r}} \sum_{l \ge 1} 
        2 i \: \sin (\eta e^{-\alpha l}) 
        \: \sin(  \eta e^{-\alpha l}  + l \theta) \\
&\simeq& i  \sqrt{\frac{2}{\pi k_r}} \left[
        \eta \: \frac{\sin \theta}{\cosh \alpha - \cos \theta} +
        \eta^2 \: \frac{\cos \theta - e^{-\alpha}}
                {\cosh 2 \alpha - \cos \theta} \right] \;. \nn
\eea
From these expressions, one can calculate the differential cross
section. The square of the first term in $f(\theta)$ gives the even
part of the cross section
\be
\label{eq:sigma_e}
\frac{d \sigma_e}{d \theta} = \frac{2\:\eta^2}{\pi k_r} \: \frac{
        \sin^2 \theta}{(\cosh \alpha - \cos \theta)^2} \;,
\ee
which contributes to a finite transport cross section but gives a
vanishing transverse cross section $\sigma_\perp$. The next order term
is odd in $\eta$ and $\theta$ 
\be
\label{eq:sigma_o}
\frac{d \sigma_o}{d \theta} = \frac{4\:\eta^3}{\pi k_r} \: 
\frac{ \sin \theta \: ( \cos \theta - \cosh 2 \alpha + \sinh \alpha )}
{(\cosh \alpha - \cos \theta) (\cosh 2 \alpha - \cos \theta)} \;,
\ee
and does contribute to $\sigma_\perp$. If one considers the limit
$\alpha \to 0$ at this point, the cross section reduces to the
expression shown in eq.\ \ref{eq:diffsig}, and the transverse cross
section would be given by eq.\ \ref{eq:sigma_perp_bb}. The problem is
that a series expansion of $d \sigma_o/d \theta$ in powers of $\alpha$
yields divergent contributions to $\sigma_\perp$ coming from small
angles. The expression above, however, is completely regular for
finite $\alpha$. One then calculates the transverse cross section
first, which yields a regular function of $\alpha$ whose series
expansion is given by
\be
\label{eq:s_p_exp}
\sigma_\perp \simeq \frac{8}{3 k_r}\:\eta^3\:\left[
        1 - \frac{\alpha}{3} 
        + \cdots
        \right] \to \frac{8 \: \eta^3}{3 k_r} 
                = - \frac{k^3 \kappa_0^3}{24 k_r c^3} \;,
\ee
in complete agreement with eq.\ \ref{eq:sigma_perp_aa}. The reason for
the apparent contradiction between eqs.\ \ref{eq:sigma_perp_aa} and 
\ref{eq:sigma_perp_bb} comes from important small-angle contributions
which were neglected in the simple-minded approach leading to eq.\
\ref{eq:sigma_perp_bb}. It is clear, however, that no contribution of
order lower than $\eta^3$ is present. 

Why does Cleary's calculation give a very different answer? In
Cleary's  \cite{cleary} and Sonin's \cite{sonin97} calculations, 
eq.\ \ref{eq:sigma_perp_partial} was set equal to
\be
\label{eq:cleary_calc}
\sigma_\perp = \frac{1}{k_r} \sum_l \sin 2 (\delta_l - \delta_{l+1})
\;,
\ee
which can be obtained from eq.\ \ref{eq:sigma_perp_partial} by adding
and subtracting the divergent series $(1/k_r)\sum \sin 2\delta_l$ and 
$(1/k_r)\sum \sin 2\delta_{l+1}$. A result linear in $\eta$ is thus 
obtained from an expression that is proportional to $\eta^3$.

The transverse force is usually calculated by inserting the transverse
cross section (eq.\ \ref{eq:s_p_exp}) into the phonon current (note
that for this calculation we have ${\bf v}_V = {\bf v}_s = 0$):
\bea
{\bf F}_t^{\rm ph} &=& \int \! 
\frac{d^3 k}{(2 \pi)^3} \: \sigma_\perp (k)
\: n( {\bf k}) \:  (\hbar c {\bf k} \times {\bf e}_z) 
\nn \\
& & 
\hspace{-1cm} =
        - ({\bf e}_z \! \times \! {\bf v}_n) 
        \frac{\hbar^2}{6 \pi^2 k_B T} \! \int \!\! dk \: 
        \sigma_\perp (k) \: \frac{k^4 e^{\hbar c k/k_B T}}
        {[e^{\hbar c k/k_B T} - 1]^2} \: . \!\!
\eea
Without going into details \cite{wexlerPHD} it is clear that the $k^2$
dependence of $\sigma_\perp$ (eq.\ \ref{eq:s_p_exp}) yields a
transverse force proportional to $T^6$ as stated in eq.\
\ref{eq:cleary'}. In contrast, the normal density is proportional to 
$T^4$ as can be clearly seen from  
\be
{\bf j}^{\rm ph} = \int \frac{d^3 k}{(2 \pi)^3} \: 
        \hbar {\bf k} \: n( {\bf k}) 
\simeq \frac{2 \pi^2}{ 45} \frac{(k_B T)^4}{\hbar^3 c^5} \: {\bf v}_n
 =  \rho_n  \: {\bf v}_n \; .
\label{eq:phonon_current}
\ee

We have seen that within this framework the result is of order
$\eta^3$, but that the coefficient of this term depends on the
limiting processes which are used.  There are other reasons that make
the result suspect.  The
first is that to the order of eq.\ \ref{eq:cleary'} there are corrections
to the simplest hydrodynamic expression for the phase shifts which might
contribute to the same order, but which cannot possibly contribute to a
lower order in $T$.
The second is that that there should be contributions from the s-wave
phase shift $\delta_0$, which will depend on the details of the vortex
core structure. It is easy to see from eq.\ \ref{eq:sigma_perp_partial}
that for $\delta_0 \neq 0$ there is a contribution to $\sigma_\perp$
given  by $(4/k_r) \: \sin 2 \eta \: \sin^2 \delta_0 \sim \kappa_0
\delta_0^2$. While usual scattering problems in two dimensions have
s-wave phase shifts that depend logarithmically on the
momentum $k$, for this problem the boundary conditions are such
\cite{fetter} that $\delta_0 \sim (k_r \: a)^2$, where $a$ is of the
order of the vortex core radius. In this case the additional
contribution to the transverse force will be of order 
\be 
  \label{eq:cleary''}
  (v_V-v_n)\kappa_0^5 T^8/\hbar^7c^{13}\;.  
\ee 

Although eq.\ \ref{eq:cleary'} might be canceled by corrections to the
hydrodynamic theory, there is no chance that eq.\ \ref{eq:cleary''}
can cancel exactly, since it is strongly dependent on the details of
the core.  

\section{Conclusions}

Our analysis has shown that a correct evaluation of Cleary's formula
\cite{cleary} for the transverse scattering of phonons by a moving
vortex gives a transverse force much smaller than the one he obtained,
and which seems to be widely accepted.  However, it is nonzero, and
slightly sensitive to the details of the vortex core, which seems to
contradict our claim \cite{TAN,wexler97} that the transverse force on an
isolated vortex is independent both of the structure of the vortex
core and of the normal fluid velocity.

This is a point that requires more detailed study, but we believe that
the Cleary calculation, involving the $\sin\theta$ weighted average of
the differential cross section, omits the most important effect of the
vortex on the phonons.  This approach assumes that the incoming wave
is asymptotically free, but actually for a system of radius $R$ each
angular momentum mode has an energy shift $c\delta_l/\pi R$, and so
this phase shift leads to a different equilibrium population of
positive and negative $l$ modes.  The shift is inversely proportional
to $R$, but the number of modes contributing is proportional to $R$,
so this effect is not obviously negligible \cite{wexlerPHD}.

The only way we know to take this into account is to use the method
which we used in Ref.\ \onlinecite{TAN}, but with more careful
examination of the boundary conditions than was given in that paper.
We are working on a careful evaluation of the problem of
noninteracting phonons in a moving vortex, and our preliminary results
suggest that the phonons give a negative contribution close to $
-\rho_n\bm{\kappa}_0\times{\bf v}_V$, which must be added to the
zero-temperature result $\rho \bm{\kappa}_0\times{\bf v}_V$.  This
term comes from the negative net circulation of the phonon momentum
around the vortex.  This results need to be checked carefully, since
it is in accord with our own preconceptions.

\acknowledgements

We would like to thank Michael Stone, Edouard Sonin, Ping Ao, Xiao-Mei
Zhu, Qian Niu, Lev Pitaevskii, Sandy Fetter, Alan Dorsey, Jian-Ming
Tang and Jean-Yves Fortin for numerous helpful discussions. 
This work was supported by the NSF through grants No. DMR-9528345 and
DMR-9628926.


%
%

\end{multicols}
\end{document}